\documentclass[runningheads]{llncs}
\pdfoutput=1
\usepackage[utf8]{inputenc} 
\usepackage[T1]{fontenc}    
\usepackage{hyperref}       
\usepackage{url}            
\usepackage{booktabs}       
\usepackage{amsfonts}       
\usepackage{nicefrac}       
\usepackage{microtype}      
\usepackage{xcolor}         
\usepackage{amsmath}
\usepackage{bm}
\usepackage{bbm}
\usepackage{todonotes}
\usepackage{graphicx}       
\usepackage{algorithm}
\usepackage{algpseudocode}
\usepackage{caption}
\usepackage{multirow}
\usepackage{makecell}

\DeclareMathOperator*{\argmin}{arg\,min}

\begin{document}
\title{GREASE: Generate Factual and Counterfactual Explanations for GNN-based Recommendations}
\titlerunning{GREASE}
%
\author{Ziheng Chen\inst{1}\orcidID{0000-0002-2585-637X} \and
Fabrizio Silvestri\inst{2}\orcidID{0000-0001-7669-9055} \and
Jia Wang\inst{3}\orcidID{0000-0002-3165-7051} \and
Yongfeng Zhang\inst{4}\orcidID{0000-0002-1243-1145} \and 
Zhenhua Huang\inst{5}\orcidID{0000-0001-8659-4062} \and
Hongshik Ahn\inst{1}\orcidID{0000-0002-8924-6159} \and
Gabriele Tolomei\inst{2}\orcidID{0000-0001-7471-6659}
}

\authorrunning{Z. Chen et al.}
%
\institute{Stony Brook University, Stony Brook, NY 11794, USA\\ 
\email{ziheng.chen@stonybrook.edu}
\and 
Sapienza University of Rome, Rome 00185, Italy\\ 
\email{\{fabrizio.silvestri,gabriele.tolomei\}@uniroma1.it}
\and
The Xi'an Jiaotong-Liverpool University, Suzhou 215000, China\\
\email{jia.wang02@xjtlu.edu.cn}
\and 
Rutgers University, Piscataway, NJ 08854, USA\\
\email{yongfeng.zhang@rutgers.edu}
\and 
South China Normal University, Guangzhou City 510631, China\\
\email{huangzhenhua@scnu.edu.cn}
}
%
\maketitle              

\newcommand{\Prob}{\mathbb{P}}
\newcommand{\R}{\mathbb{R}}
\newcommand{\Z}{\mathbb{Z}}
\newcommand{\E}{\mathbb{E}}
\newcommand{\insta}{\bm{x}}
\newcommand{\X}{X}
\newcommand{\G}{\mathcal{G}}
\newcommand{\advG}{\widetilde{\G}} 
\newcommand{\Gobs}{\G^{\text{obs}}}
\newcommand{\U}{\mathcal{U}}
\newcommand{\I}{\mathcal{I}}
\newcommand{\V}{\mathcal{V}}
\newcommand{\Vnew}{\V^{\text{new}}}
\newcommand{\Vadv}{\V^{\text{adv}}}
\newcommand{\edges}{\mathcal{E}}
\newcommand{\edgesnew}{\edges^{\text{new}}}
\newcommand{\edgesobs}{\edges^{\text{obs}}}
\newcommand{\edgesadv}{\edges^{\text{adv}}}
\newcommand{\graph}{\G=(\V,\edges)}
\newcommand{\bgraph}{\G=(\U, \I, \edges)}
\newcommand{\neigh}{\mathcal{N}}
\newcommand{\adjM}{A}
\newcommand{\advadjM}{\widetilde{A}}
\newcommand{\adjMij}{{A}_{i,j}}
\newcommand{\dataset}{\mathcal{D}}
\newcommand{\train}{\dataset_{\text{train}}}
\newcommand{\test}{\dataset_{\text{test}}}
\newcommand{\features}{\mathcal{X}}
\newcommand{\labels}{\mathcal{Y}}
\newcommand{\hypspace}{\mathcal{H}}
\newcommand{\params}{\bm{\theta}}
\newcommand{\w}{\bm{\omega}}
\newcommand{\h}{\bm{h}}
\newcommand{\advh}{\widetilde{\bm{h}}}
\newcommand{\hyp}{h_{\params}}
\newcommand{\gnn}{g(\adjM, \X; \W)}
\newcommand{\model}{h^*}
\newcommand{\loss}{\ell}
\newcommand{\ladv}{\ell_{\text{adv}}}
\newcommand{\ldist}{\ell_{\text{dist}}}
\newcommand{\lnew}{\ell_{\text{new}}}
\newcommand{\Loss}{\mathcal{L}}
\newcommand{\LF}{\Loss_{FA}}
\newcommand{\LC}{\Loss_{CF}}
\newcommand{\ind}{\mathbbm{1}}

\begin{abstract}
Recently, graph neural networks (GNNs) have been widely used to develop successful recommender systems. 
Although powerful, it is very difficult for a GNN-based recommender system to attach tangible explanations of why a specific item ends up in the list of suggestions for a given user.
Indeed, explaining GNN-based recommendations is unique, and existing GNN explanation methods are inappropriate for two reasons. 
First, traditional GNN explanation methods are designed for node, edge, or graph classification tasks rather than ranking, as in recommender systems.
Second, standard machine learning explanations are usually intended to support skilled decision-makers. 
Instead, recommendations are designed for {\em any} end-user, and thus their explanations should be provided in user-understandable ways.

In this work, we propose GREASE, a novel method for explaining the suggestions provided by {\em any} black-box GNN-based recommender system. 
Specifically, GREASE first trains a surrogate model on a target user-item pair and its $l$-hop neighborhood. 
Then, it generates both factual and counterfactual explanations by finding optimal adjacency matrix perturbations to capture the {\em sufficient} and {\em necessary} conditions for an item to be recommended, respectively.
Experimental results conducted on real-world datasets demonstrate that GREASE can generate concise and effective explanations for popular GNN-based recommender models.

\end{abstract}

\section{Introduction}
\label{sec:intro}
To alleviate the information overload caused by the tremendous amount of data that existing online services expose to end-users, many digital platforms rely on {\em recommender systems} to help their customers choose the best ``content'' via personalized suggestions.
Indeed, answering questions like ``{\em What smartphone should I buy from this e-commerce website?}'' or ``{\em What TV series should I watch on this streaming platform?}'' has become unfeasible for the user without machine-based support.

A broad spectrum of techniques for developing successful recommender systems have been proposed in the literature: most notably, {\em content-based}~\cite{pazzani2007content}, 
{\em collaborative filtering}~\cite{schafer2007collaborative,sarwar2001item}, {\em collaborative reasoning}~\cite{chen2021neural,chen2022graph}, and mixtures of those~\cite{burke2002hybrid,zhang2022neuro}.
Regardless of the approach, recommender systems have benefited from the advancement of machine learning (ML): from classical ML models~\cite{koren2009ieee,rendle2010icdm} to deep learning (DL) models based on more complex neural network (NN) architectures~\cite{guo2017ijcai,he2017sigir,he2017www}.
Amongst DL models, recommender systems based on graph neural networks (GNNs) have recently gained momentum~\cite{gao2022wsdm,he2020sigir,wang2019sigir}. 
As a matter of fact, GNNs have become a powerful technique to tackle several predictive modeling tasks in many cutting-edge domains~\cite{bastings2017acl,shi2020cvpr,wang2019sigir}.

One of the reasons why GNN-based recommender systems stand out from competitors is that they can intrinsically establish correlations amongst users, items, and associated features through information propagation on graphs derived from user-item interactions.
Specifically, GNNs learn hidden representations (i.e., embeddings) of each node (user or item) and edge (user-item preference) by repeatedly aggregating neighborhood embeddings.
This way, high-order information and multi-hop relevance among graph entities (nodes or edges) are encoded in the corresponding learned embeddings. 
Thanks to such a powerful modeling ability, GNNs have achieved outstanding performance and become state-of-the-art models for recommender systems~\cite{he2020sigir,huang2021aaai,wang2019sigir,wu2020tkde}.

Despite their effectiveness, GNN-based recommender systems lack inherent explainability, like many other DL models. 
In fact, the new challenge for any powerful recommender system would be to attach an {\em explanation} to each suggested item to clarify why {\em that} specific item has been recommended to a given user.

In the recent years, a large body of work on {\em eXplainable Artificial Intelligence} (XAI) have been proposed in the literature~\cite{tolomei2017kdd,tolomei2021tkde,mothilal2020fat,karimi2020aistats,le2020grace,lucic2019focusAAAI,siciliano2022newron}. 
Some of these studies also attempt to make GNNs more interpretable, e.g., via subgraph explorations~\cite{yuan2021icml} or counterfactual explanations~\cite{lucic2022aistats}.
However, existing explanation methods for GNNs do not easily adapt to interpret GNN-based recommendations, mostly due to two reasons.
First, traditional GNN explanation methods are designed for node, edge, or graph classification tasks.
Instead, explaining recommendations would require to figuring out the rationale behind a whole ranking of items.
Second, standard ML explanations are usually intended to support skilled decision-makers (e.g., a banker communicating why a customer has their mortgage refused by the automatic credit-risk classifier). 
On the other hand, recommendations are designed for {\em any} end-user, and thus their explanations should be provided in user-understandable ways.
These challenges make explanation methods for GNN-based recommender systems unique, and therefore we need specific approaches to address them.

In response to this demand, we propose a {\em \textbf{G}NN-based \textbf{RE}commend\textbf{A}tion \textbf{S}ystem \textbf{E}xplainer} (GREASE).
GREASE operates as follows.
First, it trains a surrogate model of a fully trained GNN on a target user-item pair and its $l$-hop neighborhood.
Then, it generates both factual and counterfactual explanations by finding optimal adjacency matrix perturbations to capture the {\em sufficient} and {\em necessary} conditions for an item to be recommended, respectively.
Intuitively, the sufficient (i.e., factual) conditions for an item to be recommended to a user are the set of user-item interactions in the $l$-hop neighborhood that allows that item to be included in the final list of top-$k$ recommended items.
An example of a factual explanation could be the following: ``{\em You have been recommended item} X {\em because you have interacted with} Y''.
On the other hand, the necessary (i.e., counterfactual) conditions are the set of user-item interactions in the $l$-hop neighborhood without which the item will {\em not} be included in the final list of top-$k$ recommended items.
An example of a counterfactual explanation could be the following: ``{\em If you had not interacted with} Y {\em then you would not have recommended item} X''.

Overall, the main contributions of our work are as follows:
\begin{itemize}
    \item[{\em (i)}] We develop GREASE, a method for generating explanations for recommendations provided by {\em any} GNN-based recommender system.
    \item[{\em (ii)}] GREASE operates in a ``black-box'' setting, i.e., it needs only to access the output predictions of the GNN-based recommender system to explain.
    \item[{\em (iii)}] GREASE can generate both factual and counterfactual explanations.
    \item[{\em (iv)}] We validate GREASE on two datasets (\textit{LastFM} and \textit{Yelp}) in combination with two famous GNN-based recommender systems (LightGCN and NGCF); results demonstrate the superiority of our method against baseline approaches.
\end{itemize}

The remainder of this paper is organized as follows. 
In Section~\ref{sec:related}, we review relevant contributions to our work proposed in the literature.
We provide some background concepts in Section~\ref{sec:background} and we formulate our problem in Section~\ref{sec:problem}.
In Section~\ref{sec:method}, we describe our proposed method GREASE, which we validate through experiments in Section~\ref{sec:experiments}. 
Finally, we conclude our work in Section~\ref{sec:conclusion}.

\section{Related Work}
\label{sec:related}
The set of relevant work to this paper cover two main areas: {\em (i) GNN-based recommender systems} and {\em (ii) explanation methods for GNNs}.
Below, we review the most significant contributions in these areas.

\noindent{{\bf {\em GNN-based recommender systems}.}}
In the past few years, many works on GNN-based recommendation have been proposed. 
Most of these contributions apply GNN to the original user-item bipartite graph directly, with a significant impact on the effectiveness and efficiency~\cite{chen2020aaai,he2020sigir,chong2019ijcai,sun2020sigir,wang2019sigir,wu2020sigir}.
Multi-GCCF~\cite{sun2019icdm} and DGCF~\cite{liu2020arxiv} propose to address the first issue by adding artificial edges between two-hop neighbors on the original graph to obtain the user-user and item-item graph. In this way, the proximity information among users and items can be explicitly incorporated into user-item interactions.

Existing GNN-based recommender systems differentiate from each other also for the node aggregation and update strategies used. 
For example, NGCF~\cite{wang2019sigir} employs element-wise product to augment the items' features the user cares about or the users' preferences for item features.
NIA-GCN~\cite{sun2020sigir} proposes the pairwise neighborhood aggregation approach to explicitly capture the interactions among neighbors. 
Inspired by the GraphSAGE~\cite{hamilton2017neurips}, some works~\cite{chong2019ijcai,sun2019icdm,rex2018kdd} adopt concatenation function with nonlinear transformation to update node and neighborhood representations.
LightGCN~\cite{he2020sigir} and LR-GCCF~\cite{chen2020aaai} simplify the update operation by removing the non-linearities, thereby retaining or even improving performance and
increasing computational efficiency.

The aggregation and update operations are applied layer by layer to generate the representations of nodes for each depth of a GNN. 
Thus, the overall representations of users and items are required for the downstream prediction task.
A mainstream approach is to use the node vector in the last layer as the final representation~\cite{chong2019ijcai,rex2018kdd}.
Recent studies, however, employ different methods to integrate messages from different layers.

A complete survey on GNN-based recommender systems can be found in~\cite{wu2022csur}.

\noindent{{\bf {\em Explanation methods for GNNs}.}}
Several GNN XAI approaches have been proposed, and a recent survey of the most relevant work is presented in~\cite{yuan2020explainability}.

The majority of existing methods provide a {\em factual} explanation in the form of a subgraph of the original graph that is deemed to be important for the prediction~\cite{baldassarre_explainability_2019,duval2021graphsvx,lin_causal_2021,luo_parameterized_2020,pope_explainability_2019,schlichtkrull_interpreting_2020,vu2020pgmexplainer,ying_gnnexplainer_2019,yuan2021icml}.
Such methods are analogous to popular XAI methods like LIME~\cite{ribeiro2016lime} or SHAP~\cite{lundberg_unified_2017}, which identify relevant features for a particular prediction for tabular, image, or text data. 
All of these methods require the user to specify the size of the explanation in advance, i.e., the number of features (or edges) to keep.
More recently, Lucic et al.~\cite{lucic2022aistats} have introduced CF-GNNExplainer. This is the first method to extract {\em counterfactual} explanations from GNN model predictions.
It does so by automatically finding the optimal perturbation of the input graph that results in a prediction switch.

Regardless of the approach, however, existing explanation methods for GNNs are designed for node, edge, or graph classification tasks.
Instead, explaining GNN-based recommendations would require to figuring out the rationale behind a whole ranking of items, rather than a single prediction.
Combining the work of Tan et al.~\cite{tan2021counterfactual,tan2022learning}, our method (GREASE) is different from other existing GNN XAI solutions as it provides a general framework to extract {\em both} factual and conterfactual explanations for GNN-based recommender systems.

\section{Background and Notation}
\label{sec:background}
We consider the standard collaborative filtering recommendation task.
Specifically, let $\mathcal{U}=\{u_1,u_2,\ldots,u_m\}$ be a set of $m$ users, and $\mathcal{I}=\{i_1,i_2,\ldots,i_n\}$ be a set of $n$ items.
We represent the interactions between users and items with a binary user-item matrix $Y\in\{0,1\}^{m\times n}$, where $Y_{u,i}=y_{u,i} = 1$ indicates that user $u$ has interacted with item $i$, or $0$ otherwise.
Notice that this resembles the case where user's preference on items is expressed via implicit feedback (e.g., clicks/likes) rather than explicit relevance judgements (e.g., ratings).
Either way, the recommendation problem can be formulated as estimating the value $\hat{y}_{u,i}$ for each $u\in \mathcal{U}$ and each $i\in \mathcal{I}$ as follows:
\[
\hat{y}_{u,i} = f(\h_u, \h_i),
\]
where $\h_u,\h_i\in \R^d$ are suitable user and item representations, respectively, and $f: \R^d \times \R^d \mapsto [0,1]$ is a function that measures the preference
score for user $u$ on item $i$.\footnote{A similar reasoning would apply if we instead considered explicit ratings, i.e., $f: \R^d \times \R^d \mapsto \R$.}
The predicted score computed with $f$ is in turn used to rank items to populate a list $\I^k_{u}\subseteq \I$ of the top-$k$ recommended items for user $u$, i.e., the list of $k$ items most likely relevant to user $u$ according to $f$.

In this work, we focus on GNN-based recommender systems, namely we assume that both user and item latent representations (i.e., the embeddings $\h_u$ and $\h_i$) are learned through a GNN $g$.
Concretely, we consider the bipartite graph $\bgraph$ obtained from user-item interactions represented by $Y$, where $\U$ and $\I$ correspond to two disjoint and independent sets of nodes and $\edges \subseteq \U \times \I$ is the set of edges. 
There exists an edge connecting a node $u\in \U$ to another node $i\in \I$ iff user $u$ interacts with item $i$.
Thus, the topology of $\G$ is encoded by its adjacency matrix $\adjM \in \lbrace 0,1 \rbrace^{m \times n}$, where $\adjM_{u,i} = y_{u,i}$, i.e., $\adjM_{u,i} = 1$  iff $(u, i) \in \edges$. 
We refer to the overall set of nodes of $\G$ as $\V = \U \cup \I$. 
Moreover, each node $v\in \V$ can be also associated with a $p$-dimensional real-valued feature vector, which represents one row of the feature matrix $\X \in \mathbb{R}^{(m+n) \times p}$.

The embedding of each node is learned through $g$ by iteratively updating the node's features based on the neighbors' features.
Formally, let $\h^l_v$ denote the embedding of the generic node $v \in \V$ at the $l$-th layer of $g$. 
Hence,
\[
    \h^l_v = g(\adjM^l_v,\X^l_v;\params_g) = \phi(\h^{l-1}_v, \psi(\{\h^{l-1}_w~|~w\in \neigh(v)\})),
\]
where: $\neigh(v) = \{w\in \V~|~(v,w)\in \edges\}$ is the $1$-hop neighbors of $v$, i.e., the set of all nodes that are adjacent to $v$; $\adjM^l_v$ and $\X^l_v$ are the adjacency matrix and the feature matrix of nodes in the subgraph $\G^l_v$ of $\G$, induced by $v$ and its $l$-hop neighbors; $\phi$ and $\psi$ are arbitrary differentiable functions (i.e., neural networks): $\psi$ is a permutation-invariant operator that {\em aggregates} the information from the $l$-hop neighborhood of $v$; $\phi$ {\em updates} the node embedding of $v$ by combining information from the previous layer; $\params_g$ are the trainable parameters of $g$.
As it turns out, the number of hidden layers in $g$ determines the set of neighbors included while learning each node's embedding.
Finally, let $\h^l_v = g(\adjM^l_v,\X^l_v;\params_g)$ be the embedding of the generic node $v\in \V$.
Thus, we assume that the predicted rating score of user $u$ for the item $i$ is approximated with $f$, defined as follows:
\[
f(\h^l_u, \h^l_i; \params_f) = f (g(\adjM^l_{u,i},\X^l_u;\params_g), g(\adjM^l_{u,i},\X^l_i;\params_g); \params_f),
\]
where we denote by $\adjM^l_{u,i}$ the adjacency matrix of the subgraph $\G^l_{u,i} = \G^l_u + \G^l_i$ induced by the union of the $l$-hop neighborhood of both user $u$ and item $i$.
From now on, however, we will consider $l$ fixed and omit the corresponding superscript, unless otherwise needed.
Notice that $f$ can be any function parameterized by $\params_f$ that takes as input two node embeddings learned by $g$ and outputs the corresponding preference score, e.g., $f$ can be itself another neural network like a multilayer perceptron. 
Otherwise, $f$ can simply compute the dot product or the cosine similarity between the two input node embeddings. 

\section{Problem Formulation}
\label{sec:problem}
Without loss of generality, we consider a well-trained GNN-based recommender system $f$, which computes the preference score of a user $u$ on an item $i$ as the dot product of the node embeddings learned through an $l$-layer GNN $g$, namely:
\[
 \hat{y}_{u,i} = f(g(A_{u,i},X_{u}), g(A_{u,i},X_i))  = f(\h_u, \h_i) = \h^{T}_u \h_i.
\]
Let $\I^k_{u}$ be the list of top-$k$ recommended items for the user $u$.
Informally, our goal is to attach an {\em explanation} to every item $i \in \I^k_{u}$ that clarifies why $i$ has been recommended to $u$.

To achieve this goal, we envision two methods that derive from two opposite yet complementary approaches: {\em factual} ($FA$) and {\em counterfactual} ($CF$) reasoning.
Specifically, factual reasoning seeks a subgraph $\G^{FA}$ from the user-item interaction graph $\G_{u,i}$ whose information is {\em sufficient} to include item $i$ into user $u$'s top-$k$ list $\I^k_{u}$. 
On the other hand, counterfactual reasoning targets on a subgraph $\G^{CF}$ whose information is {\em necessary}, i.e., without which item $i$ would {\em not} be in the top-$k$ list of recommendations $\I^k_{u}$. 

Extracting two subgraphs $\G^{FA}$ and $\G^{CF}$ from $\G_{u,i}$ is equivalent to find two adjacency matrices $A^{FA}$ and $A^{CF}$ in place of the original $A_{u,i}$.
Hence, the goal of factual and counterfactual reasoning boils down to learning two edge mask matrices $P^{FA}$ and $P^{CF}$, such that $A^{FA} = A_{u,i} \odot P^{FA}$ and $A^{CF} = A_{u,i} \odot P^{CF}$, where $\odot$ indicates the Hadamard (i.e., element-wise) matrix product.

Formally, for factual explanations we perturb the $l$-hop neighborhood of both $u$ and $i$, so that $\h_u \leadsto \h^{FA}_u$ and $\h_i \leadsto \h^{FA}_i$, where:
\[
\h^{FA}_{u}=g(A^{FA}, X_{u}),\quad \h^{FA}_{i}=g(A^{FA}, X_{i}),
\]
and $A^{FA} = A_{u,i} \odot P^{FA}$.
Eventually, this will lead to a new prediction score $\hat{y}^{FA}_{u,i} = f(\h^{FA}_{u}, \h^{FA}_{i})$.
In fact, notice that, because of the neighborhood aggregation operations performed by the GNN $g$, all the embeddings of the nodes in $\G_{u,i}$ will change if $A_{u,i}$ is modified.
Therefore, in general, we may obtain a new prediction score $\hat{y}^{FA}_{u,j}$ for every $j\in \I^k_u$ that was also in the $l$-hop neighborhood.

However, instead of directly optimizing the rank of the target item $i$ we want to explain, factual reasoning seeks to maximize its predicted rating score $\hat{y}^{FA}_{u,i}$ until $i$ is recommended, i.e., until $i$ is pushed in the list of top-$k$ recommended items $\I^k_u$, regardless of its rank. 
Thus, we define the following factual explanation loss function for a fixed user-item pair $(u,i)$:
\begin{equation}
\label{eq:fa-loss}
\LF(P^{FA})=-\hat{y}^{FA}_{u,i} \cdot \ind_{\I_u^k}(i),
\end{equation}
where $\ind$ is the well-known $0$-$1$ indicator function: $\ind_{\I_u^k}(i)$ will be $1$ if $\I_u^k$ contains item $i$, $0$ otherwise. $\LF$ is minimized when item $i$ appears in the list of top-$k$ recommendations $\I^k_u$.
Similarly, for counterfactual explanations, we perturb the $l$-hop neighborhood of both $u$ and $i$, so that  $\h_u \leadsto \h^{CF}_u$ and $\h_i \leadsto \h^{CF}_i$, where:
\[
\h^{CF}_{u}=g(A^{CF},X_{u}),\quad \h^{CF}_{i}=g(A^{CF},X_{i}),
\]
and $A^{CF} = A_{u,i}\odot P^{CF}$.
As opposed to factual reasoning, though, the goal of counterfactual explanations is to directly minimize the new prediction score $\hat{y}^{CF}_{u,i} = f(\h^{CF}_u, \h^{CF}_i)$ until $i$ is {\em no longer} recommended, i.e., to evict item $i$ from the list of top-$k$ recommended items $\I^k_u$. 
Formally, we define the counterfactual loss function for a fixed user-item pair $(u,i)$ as follows:
\begin{equation}
\label{eq:cf-loss}
\LC(P^{CF})=\hat{y}^{CF}_{u,i} \cdot \ind_{I^k_u}(i),
\end{equation}
where $\ind$ is the well-known $0$-$1$ indicator function, and $\LC$ is minimized when item $i$ is removed from the list of top-$k$ recommendations $\I^k_u$.

Notice that both factual and counterfactual loss incorporate a $0$-$1$ indicator function term, which is non-convex and non-smooth, thus hard to optimize using gradient-based methods. 
In practice, we replace $\ind_{I^k_u}(i)$ with the following term:
\[
\text{ReLU}\Big(\hat{y}_{u,i} - \min\big\{\bigcup_{j\in I^k_u}\hat{y}_{u,j}\big\} + \varepsilon \Big) = \max\Big(0, \hat{y}_{u,i} - \min\big\{\bigcup_{j\in I^k_u}\hat{y}_{u,j}\big\} + \varepsilon \Big),
\]
which evaluates to $0$ for any item $i$ whose predicted preference score is smaller than that of the last ranked item in the top-$k$ list, or is greater than $0$ otherwise. 
The small tolerance $\varepsilon > 0$ guarantees that the input of the ReLU function is positive until considering the last $k$-th item of the list, negative otherwise. 
In this work, we set $\varepsilon=0.05$.

Furthermore, we want the selected subgraphs $\G^{FA}$ and $\G^{CF}$ to be as sparse as possible. Hence, we generate factual and counterfactual explanations by minimizing a joint loss function of the form:
\begin{equation}
\label{eq:loss}
\Loss(P) = \Loss_{\text{exp}}(P) + \beta \Loss_{\text{dist}}(P),
\end{equation}
where $P$ is a generic perturbation matrix (i.e., either $P^{FA}$ or $P^{CF}$) and $\Loss_{\text{exp}}$ works as a placeholder for explanation loss, which could be substituted by $\LF$ or $\LC$ for generating factual and counterfactual explanations, respectively. 
Also, $\Loss_{\text{dist}}$ measures the distance between the perturbed matrix $A_{u,i}\odot P$ and the original adjacency matrix $A$. 
For example, $\Loss_{\text{dist}}(P) =||A_{u,i}\odot P-A_{u,i}||_p$ computes the $L^p$-norm of the difference between the two matrices. In this work, we set $\Loss_{\text{dist}}(P) =||A_{u,i}\odot P-A_{u,i}||_1$, i.e., the $L^1$-norm.

Our problem can be thus formulated as the following optimization task:
\begin{equation}
\label{eq:objective}
P^* = \argmin_{P} \Loss(P).
\end{equation}
Eventually, the best factual and counterfactual explanation is found by computing $A^{FA^*} = A_{u,i} \odot P^{FA^*}$ and $A^{CF^*} = A_{u,i} \odot P^{CF^*}$, respectively.

\section{Proposed Method: GREASE}
\label{sec:method}
We propose GREASE to solve the objective defined in Equation~\ref{eq:objective}.
Theoretically, to find the optimal perturbation matrix $P^*$, we should take the gradient w.r.t. $P$ of the joint loss function $\Loss(P)$ defined in Equation~\ref{eq:loss}. 
However, since GREASE operates in a black-box setting, the gradient of the model $g$ (and therefore $f$) is unknown. 
This assumption is restrictive yet pretty standard, especially in a commercial scenario, where the recommender system's internals (e.g., the model's architecture, parameters, or gradient) are rarely disclosed.
Therefore, we first learn a local approximator $\widetilde{g}$.
This surrogate model tries to resemble $g$ in learning the embeddings of each node in the subgraph $\G^l_{u,i}=\G^l_u +\G^l_i$ induced by the union of the $l$-hop neighborhood of both user $u$ and item $i$.
The intuition is that the aggregation mechanism behind node embeddings is complex over the complete network. But in the local neighborhood of a target user-item pair, such aggregation process can be relatively simple, hence amenable to be extracted by a simpler model. 
Then, in our objective function, we replace $g$ with a surrogate approximation of it $\widetilde{g}$, and find the optimal $P$ via gradient-based optimization. 

\subsection{Train a Surrogate GNN Model}
Given a trained black-box GNN model $g$, a user-item pair $(u,i)$, and the subgraph $\G^l_{u,i}$ induced by the $l$-hop neighborhood of $u$ and $i$, we first learn a surrogate model $\widetilde{g}\approx g$ to approximate the embedding of each node in $\G^l_{u,i}$.
We can measure the surrogate loss function via standard mean squared error:
\[
\Loss_{\text{sur}}(g,g')=\sum_{k=1}^{|\V^{l}_{u,i}|}||g(A^{l}_{u,i},X^{l}_{k})-g'(A^{l}_{u,i},X^{l}_{k})||^{2},
\]
where $\V^l_{u,i}$, $X^{l}_{k}$, $A^{l}_{u,i}$ are the set of nodes, feature and adjacency matrix of $\G^l_{u,i}$. Both $g$ and $g'$ output a $d$-dimensional embedding vector, and the optimal approximate model $\widetilde{g}$ is found by minimizing the surrogate loss function above:
\[
\widetilde{g} = \argmin_{g'}\Loss_{\text{sur}}(g,g').
\]
Considering we aim to extract the linkage information from a heterogeneous network with two types of nodes (i.e., users and items), we adopt a Relational Graph Convolutional Network (R-GCN)~\cite{thanapalasingam2021relational}\cite{schlichtkrull2018modeling} to learn the surrogate model. 
R-GCN extends GCN to heterogeneous graphs by considering the direction and type of edges, separately. 
To allow R-GCN to pass messages through undirected edges linking users and items, we transform our original bipartite graph $\G$ by substituting each edge pair $(u,i)\in \edges$ with two pairs of symmetric edges $(u,i)$ and $(i,u)$, respectively.
Thus, the user embedding $\h_{u}$ and the item embedding $\h_{i}$ will be propagated with different weights in the generic $l$-th layer:
\[\h_{u}^{l}=\sigma\left(\sum_{i\in \mathcal{N}(u)}\frac{1}{\mathcal{N}(u)}W_{(i,u)}^{(l-1)}\h_{i}^{(l-1)}+W_{0}^{(l-1)}\h_{u}^{(l-1)}\right)\]
\[\h_{i}^{l}=\sigma\left(\sum_{u\in \mathcal{N}(i)}\frac{1}{\mathcal{N}(i)}W_{(u,i)}^{(l-1)}\h_{u}^{(l-1)}+W_{0}^{(l-1)}\h_{i}^{(l)}\right)\]
In this way, the weight matrices $W_{(i,u)}^{(l-1)}$ and $W_{(u,i)}^{(l-1)}$ are leveraged to update the users' and items' information, separately.

\subsection{Generate Explanations}
After we find the best surrogate GNN model $\widetilde{g}$, we can solve the optimization task defined in Equation~\ref{eq:objective}.
We let $\widetilde{\Loss}(P)$ denote the same loss function as the one defined in Equation~\ref{eq:objective} yet depending on $\widetilde{g}$ rather than the original GNN $g$.

In Algorithm~\ref{alg:grease}, we describe how GREASE generates both factual {\em or} counterfactual explanations for a given recommended item $i$ to a user $u$.
\begin{algorithm}[ht!]
\caption{{GREASE}}
\small
\label{alg:grease}
\begin{algorithmic}[1]
\State
{\textbf{Input:} Surrogate GNN model $\widetilde{g}$; user $u$; item $i$; feature matrix $X$; adjacency matrix $A^{l}_{u,i}$ of the subgraph $\G^l_{u,i}$ induced by the union of the $l$-hop neighborhoods of $u$ and $i$; number of iterations $M$; trade-off $\beta$; learning rate $\eta$; size $k$ of the list of recommended items $\I^k_u$; explanation flag $exp$ ({\tt FA}=factual; {\tt CF}=counterfactual)}
\State{\textbf{Output:} The factual/counterfactual explanation matrix $A^*$ associated with the optimal perturbation matrix $P^*$, or $\bot$ if no valid explanation is found.}
\item[]
\Procedure{{\tt GREASE}}{$\widetilde{g}, u, i, X, A^l_{u,i}, M, \beta, \eta, k, exp$}:
\State{$\hat{J} = J;$}
\If{$exp$ == {\tt FA}}
\State{$\hat{J}[:,u] = {\bf 0}; \quad \hat{J}[:,i] = {\bf 0};$}
\State{$P=\hat{J}; \quad \Loss_{\text{exp}}(P) = \LF(P);$}
\EndIf
\If{$exp$ == {\tt CF}} 
\State{$P=J; \quad \Loss_{\text{exp}}(P) = \LC(P);$}
\EndIf
\State{$A^{0}=A^{l}_{u,i}\odot J;$}
\State{$\hat{P}^1 = P;\quad P^* = \bot;$}
\State{$min\_loss\_dist = \infty;$}
\For{$j \in [1\ldots M]$}
\State{$P^j = \sigma(\hat{P}^j, 0.5);$}
\State{$A^j = A^l_{u,i} \odot P^j;$}
\State{$\Loss_{\text{dist}}(\hat{P}^j)=||A^j-A^{j-1}||_{1};$}
\State{$curr\_loss\_dist = \Loss_{\text{dist}}(\hat{P}^j);$}
\State {$r = f(\widetilde{g}(A^j, X_{u}),\widetilde{g}(A^j, X_{i}));$}
\If{$(exp==\texttt{FA} {\textbf{ and }} r\leq k){\textbf{ or }}(exp==\texttt{CF} {\textbf{ and }} r > k)$}
\If{$curr\_loss\_dist < min\_loss\_dist$}
\State{$P^{*}=P^j;$}
\State{$min\_loss\_dist = curr\_loss\_dist;$}
\EndIf
\EndIf
\State{
$\widetilde{\Loss}(\hat{P}^j) = \Loss_{\text{exp}}(\hat{P}^j) + \beta \Loss_{\text{dist}}(\hat{P}^j);$
}
\State $\hat{P}^{j+1} = \hat{P}^{j}-\eta \nabla_{\hat{P}^j}\widetilde{\Loss}(\hat{P}^j);$
\EndFor
\If{$P^* \neq \bot$}
\State \Return $A^{*}=A^{l}_{u,i}\odot P^{*};$
\Else 
\State \Return $\bot;$
\EndIf
\EndProcedure
\end{algorithmic}
\end{algorithm}
The algorithm returns the adjacency matrix (i.e., the subgraph) corresponding to the factual or counterfactual explanations (i.e., $A^{FA}$ or $A^{CF}$, respectively), or $\bot$ if no valid explanation is found.
To achieve that, we consider the two types of explanations, separately.
First, we create two $(m+n)\times (m+n)$ matrices of all ones: $J$ and $\hat{J}$, respectively (line 4).
Then, depending on the value of the $exp$ flag, we perform a different sequence of initialization steps.
Specifically, if we want to extract {\em factual} explanations (i.e., $exp=\texttt{FA}$), we need to start by evicting all the existing edges associated with user $u$ and item $i$.
Concretely, we nullify the $u$-th and $i$-th column of the matrix $\hat{J}$, and we assign it to the initial perturbation matrix $P$. Also, we instantiate the explanation loss to the factual explanation loss defined in Equation~\ref{eq:fa-loss} (lines 5--8).
Instead, if we want to extract {\em counterfactual} explanations (i.e., $exp=\texttt{CF}$), we simply initialize the perturbation matrix $P$ with the matrix of all ones $J$, so that at the beginning we to retain all the edges of the original graph. In addition, we set the explanation loss to the counterfactual explanation loss of Equation~\ref{eq:cf-loss} (lines 9--11).
Still, as part of the common initialization procedure, we set the current explanation adjacency matrix $A^0$ to the original adjacency matrix associated with $\G^l_{u,i}$ (line 12). Furthermore, to transform our optimization problem from discrete to continuous, we use a smoothed, real-valued perturbation matrix $\hat{P}^1$, initially set to $P$. At the beginning, the optimal perturbation matrix is undefined ($\bot$) and the minimal distance between the original and the perturbed adjacency matrices is set to infinity (lines 12--14).

The core iterative procedure of GREASE is encapsulated in the for loop between lines 15 and 29, which is repeated $M$ times.
At each iteration $j$, the first step is to binarize the current real-valued perturbation matrix. To do so, every element of $\hat{P}^j$ is passed through a sigmoid filter $\sigma$ that squashes them into a real value in the range $[0,1]$. Then, we further discretize the matrix entries using a threshold equal to $0.5$, i.e., if the entry is greater than or equal to $0.5$ it is mapped to $1$, or $0$ otherwise (line 16).
Once the perturbation matrix is binarized, with all its entries being $0$ or $1$, we compute the Hadamard product between the original adjacency matrix and the current perturbation, thereby originating a new explanation matrix $A^j$ for this iteration (line 17).
We thus compute and store the updated $L^1$-norm of the distance between the original and the current explanation matrix (lines 18--19).
Afterwards, we compute the rank $r$ of item $i$ for user $u$ according to our recommender system $f$ based on the surrogate GNN model $\widetilde{g}$ and the currently perturbed adjacency matrix $A^j$ (line 20).
We therefore check if the explanation goal is met. Specifically, for factual explanations, the computed rank must be high enough to let item $i$ be added to the top-$k$ recommendations. Instead, for counterfactual explanations, the computed rank must be low enough to let item $i$ be removed from the top-$k$ recommendations (line 21). 
In the case the explanation goal is met, we must check whether the current perturbation matrix is closer to the original adjacency matrix than the optimal perturbation computed so far. 
If so, we must then update the optimal perturbation accordingly and keep track of the new best distance (lines 21--26).
Nevertheless, we update the overall loss function as defined in Equation~\ref{eq:loss} and take one step of gradient descent to update the perturbation matrix $\hat{P}^{j+1}$, which will be considered at the next iteration $j+1$ (lines 27--28).

Eventually, if the optimal perturbation matrix is changed during the $M$ iterations, GREASE will return the final explanation matrix, i.e., either $A^{FA^*}$ or $A^{CF^*}$ (lines 30--31). Otherwise, no valid explanation is found for the given input user-item pair (line 33).

\section{Experiments}
\label{sec:experiments}
In this section, we describe the experiments conducted to validate the ability of GREASE to explain suggestions provided by GNN-based recommender systems.
\noindent {\bf \em Datasets.} 
Before we can apply GREASE, we train our GNN-based recommender systems on the following two datasets: \textit{LastFM}~\cite{lastfm-ds} and \textit{Yelp}~\cite{yelp-ds}.
An overview of the main properties of these data collections is shown in Table~\ref{tab:datasets}.
\begin{table}[ht]
\centering
\begin{tabular}{|l|c|c|c|}
\hline
\textbf{Dataset}        & \textbf{N. of Users} & \textbf{N. of Items} & \textbf{N. of Interactions}  \\ \hline
\textit{LastFM}  & 1,801 & 7,432 & 76,693 \\ \hline
\textit{Yelp}  & 31,668 & 38,048 & 1,561,406 \\ \hline
\end{tabular}
\caption{\label{tab:datasets}Main characteristics of the public datasets used.
}
\end{table}
\vspace{-10mm}

\noindent {\bf \em GNN-based Recommender Systems.} 
For both datasets, we randomly select $80\%$ of historical interactions of each user to build the training set, and treat the remaining as the test set. 
Thus, we train two different GNN-based recommender systems on each dataset, which in turn we will explain with GREASE: LightGCN~\cite{he2020sigir} and NGCF~\cite{wang2019sigir}. 
We set the number of layers of the GNN to $3$, namely the embeddings are learned from the $3$-hop neighborhood of each node. 
In this way, each user's embedding will be influenced also by the items rated by other users, and each item's embedding will be influenced also by the users who have rated other items.
Both LightGCN and NGCF are trained by minimizing binary cross-entropy loss via Adam optimizer with learning rate equal to $0.001$.
Eventually, both GNN-based recommender systems use the dot product between the node embeddings learned either by LightGCN or NGCF to compute the preference score of each user-item pair.

\noindent {\bf \em Parameter Settings.}
For each user $u$ in every dataset, we first generate the top-$k$ list or recommended items $\I^k_u$ according to the GNN-based recommender systems considered (i.e., LightGCN and NGCF), where $k$ is selected from $\{10,15,20\}$. 
Then, we consider every item $i\in \I^k_u$ and let GREASE generate a factual and a counterfactual explanation, separately. 
It is worth remarking that GREASE operates on the subgraph $\G^l_{u,i}$ induced by the $l$-hop neighborhood of $u$ and $i$ to generate such explanations. 
In this work, we set $l=2$ for all settings. 
Besides, a two-layer R-GCN is used for training the surrogate models $\widetilde{g}$ associated with the original GNN models, i.e., LightGCN and NGCF. 
The hidden dimensions are set to $32$ for \textit{LastFM} and $64$ for \textit{Yelp} datasets, respectively. 
In addition, we set $M=200$ and $\eta=0.01$ for the number of iterations and the learning rate of GREASE (see Algorithm~\ref{alg:grease}). 
The trade-off $\beta$ between explanation ($\Loss_{\text{exp}}$) and distance loss ($\Loss_{\text{dist}}$) is set to $1/200$ both for \textit{LastFM} and \textit{Yelp}.

\noindent {\bf \em Evaluation Metrics.}
We evaluate the quality of generated explanations according to the following three metrics: {\em Probability of Sufficiency} (PS), {\em Probability of Necessity} (PN) and {\em Explanation Cost} (EC). PS measures the percentage of factual explanations that successfully push the item $i$ into the $u$'s top-$k$ list. Similarly, PN measures the percentage of counterfactual explanations that evict $i$ from $u$'s top-$k$ list. For both PS and PN, the higher the better. EC, instead, indicates the number of edges that needs to be added or deleted according to the factual explanation and counterfactual explanation, respectively. 
To deal with the huge number of user-item pairs to be explained, we randomly sample $10\%$ users in each setting, and explain all the items in their top-$k$ list of recommendations ($k\in \{10,15,20\})$. Moreover, experiments were repeated five times (i.e., five different samples of users are randomly selected) and results are expressed as the mean $\pm$ standard deviation.

\noindent {\bf \em Baselines.}
To evaluate our algorithm, we compare it against two baselines: {\sc Random} and {\sc PersonalRank}. 
The former is meant as a sanity check, as it randomly adds/deletes edges from $\G^l_{u,i}$ to generate factual/counterfactual explanations. 
In {\sc PersonalRank}, instead, we start by modifying all the existing edges associated with nodes that have the largest PersonalRank~\cite{haveliwala2003tkde} score until we possibly get either a factual or a counterfactual explanation.\footnote{PersonalRank is the personalized version of the well-known PageRank algorithm.}

To control the complexity (i.e., EC), we limit the maximum cost for constructing factual and counterfactual explanations.
Specifically, after testing with several combination of values, we allow at most $6$ and $10$ edges to be added/removed, respectively, as these lead GREASE to the best results.

\noindent {\bf \em Results.}
In Table~\ref{tab:top-10-results}, \ref{tab:top-15-results}, and \ref{tab:top-20-results}, we compare the quality of explanations generated by our method (GREASE) with those of the two baselines ({\sc Random} and {\sc PersonalRank}), when $k=10$, $15$, and $20$, respectively.
We can observe that GREASE consistently outperforms other competitors on all the three metrics. 
Specifically, compared with {\sc PersonalRank}, GREASE improves the validity of factual explanations for the recommendations generated on {\em LastFM} by up to $15\%$ ($+12\%$ on {\em Yelp}).
Moreover, the validity of counterfactual explanations on {\em LastFM} is $21\%$ higher than those provided by {\sc PersonalRank} ($+20\%$ on {\em Yelp}). 
\begin{table*}[ht]
\scalebox{0.85}{
\centering
\begin{tabular}{lll|cc|}
\cline{4-5}
& & & \multicolumn{2}{c|}{{\bf Top-$10$ recommendations}}
\\
\hline
\multicolumn{1}{|l|}{\bf Dataset} & \multicolumn{1}{l|}{\thead{Recommendation\\Model}} & \multicolumn{1}{l|}{\thead{Explanation\\Method}} & \multicolumn{1}{c|}{\bf PN($\uparrow$)/EC($\downarrow$)} & \multicolumn{1}{c|}{\bf PS($\uparrow$)/EC($\downarrow$)}
\\
\hline
\multicolumn{1}{|l|}{} & 
\multicolumn{1}{l|}{} & 
\multicolumn{1}{l|}{GREASE} & 
\multicolumn{1}{c|}{$\mathbf{0.89} (\pm 3.00)$/$\mathbf{3.81} (\pm 0.27)$} & \multicolumn{1}{c|}{$\mathbf{0.69} (\pm 3.00)$/$\mathbf{6.21} (\pm 0.76)$}
\\
\cline{3-5}
\multicolumn{1}{|l|}{} &
\multicolumn{1}{l|}{LightGCN} & 
\multicolumn{1}{l|}{{\sc PersonalRank}} & \multicolumn{1}{c|}{$0.79 (\pm 2.00)$/$4.63 (\pm 2.50)$} & \multicolumn{1}{c|}{$0.58 (\pm 2.00)$/$6.68 (\pm 0.36)$}
\\
\cline{3-5}
\multicolumn{1}{|l|}{} &
\multicolumn{1}{l|}{} & 
\multicolumn{1}{l|}{{\sc Random}} &
\multicolumn{1}{c|}{$0.19 (\pm 3.00)$/$5.00 (\pm 1.20)$} & \multicolumn{1}{c|}{$0.11 (\pm 4.00)$/$8.50 (\pm 0.41)$}
\\
\cline{2-5}
\multicolumn{1}{|l|}{\textit{LastFM}} & 
\multicolumn{1}{l|}{} & 
\multicolumn{1}{l|}{GREASE} & 
\multicolumn{1}{c|}{$\mathbf{0.90} (\pm 6.00)$/$\mathbf{3.23} (\pm 0.33)$} & \multicolumn{1}{c|}{$\mathbf{0.70} (\pm 6.00)$/$\mathbf{6.30} (\pm 0.82)$}
\\
\cline{3-5}
\multicolumn{1}{|l|}{} &
\multicolumn{1}{l|}{NGCF} & 
\multicolumn{1}{l|}{{\sc PersonalRank}} & \multicolumn{1}{c|}{$0.79 (\pm 2.00)$/$4.63 (\pm 2.50)$} & \multicolumn{1}{c|}{$0.58 (\pm 2.00)$/$6.68 (\pm 0.36)$}
\\
\cline{3-5}
\multicolumn{1}{|l|}{} &
\multicolumn{1}{l|}{} & 
\multicolumn{1}{l|}{{\sc Random}} &
\multicolumn{1}{c|}{$0.19 (\pm 3.00)$/$5.00 (\pm 1.20)$} & \multicolumn{1}{c|}{$0.11 (\pm 4.00)$/$8.50 (\pm 0.41)$} 
\\
\hline
\hline
\multicolumn{1}{|l|}{} & 
\multicolumn{1}{l|}{} & 
\multicolumn{1}{l|}{GREASE} &
\multicolumn{1}{c|}{$\mathbf{0.91} (\pm 4.00)$/$\mathbf{3.26} (\pm 0.31)$} & \multicolumn{1}{c|}{$\mathbf{0.70} (\pm 3.37)$/$\mathbf{6.37} (\pm 0.38)$} 
\\
\cline{3-5}
\multicolumn{1}{|l|}{} & 
\multicolumn{1}{l|}{LightGCN} & 
\multicolumn{1}{l|}{{\sc PersonalRank}} & \multicolumn{1}{c|}{$0.82 (\pm 3.50)$/$4.79 (\pm 0.36)$} & \multicolumn{1}{c|}{$0.61 (\pm 2.95)$/$6.93 (\pm 0.41)$} 
\\
\cline{3-5}
\multicolumn{1}{|l|}{} & 
\multicolumn{1}{l|}{} & 
\multicolumn{1}{l|}{{\sc Random}} & 
\multicolumn{1}{c|}{$0.21 (\pm 5.30)$/$5.39 (\pm 1.32)$} & \multicolumn{1}{c|}{$0.16 (\pm 2.80)$/$7.35 (\pm 0.13)$} 
\\
\cline{2-5}
\multicolumn{1}{|l|}{\textit{Yelp}} & 
\multicolumn{1}{l|}{} & 
\multicolumn{1}{l|}{GREASE} &
\multicolumn{1}{c|}{$\mathbf{0.93} (\pm 6.00)$/$\mathbf{5.70} (\pm 0.45)$} & \multicolumn{1}{c|}{$\mathbf{0.71} (\pm 5.00)$/$\mathbf{6.21} (\pm 0.82)$} 
\\
\cline{3-5}
\multicolumn{1}{|l|}{} & 
\multicolumn{1}{l|}{NGCF} & 
\multicolumn{1}{l|}{{\sc PersonalRank}} & \multicolumn{1}{c|}{$0.83 (\pm 3.00)$/$6.22 (\pm 0.41)$} & \multicolumn{1}{c|}{$0.59 (\pm 3.00)$/$7.21 (\pm 0.38)$} 
\\
\cline{3-5}
\multicolumn{1}{|l|}{} & 
\multicolumn{1}{l|}{} & 
\multicolumn{1}{l|}{{\sc Random}} & 
\multicolumn{1}{c|}{$0.22 (\pm 3.00)$/$8.20 (\pm 1.47)$} & \multicolumn{1}{c|}{$0.15 (\pm 4.00)$/$8.60 (\pm 0.46)$}
\\
\hline
\end{tabular}
}
\caption{\label{tab:top-10-results}
PN, PS, and EC of factual and counterfactual explanations generated by GREASE, {\sc PersonalRank}, and {\sc Random} for top-$10$ recommended items.}
\end{table*}

\begin{table*}[ht]
\scalebox{0.85}{
\centering
\begin{tabular}{lll|cc|}
\cline{4-5}
& & & \multicolumn{2}{c|}{{\bf Top-$15$ recommendations}}
\\
\hline
\multicolumn{1}{|l|}{\bf Dataset} & \multicolumn{1}{l|}{\thead{Recommendation\\Model}} & \multicolumn{1}{l|}{\thead{Explanation\\Method}} & \multicolumn{1}{c|}{\bf PN($\uparrow$)/EC($\downarrow$)} & \multicolumn{1}{c|}{\bf PS($\uparrow$)/EC($\downarrow$)}
\\
\hline
\multicolumn{1}{|l|}{} & 
\multicolumn{1}{l|}{} & 
\multicolumn{1}{l|}{GREASE} & 
\multicolumn{1}{c|}{$\mathbf{0.87} (\pm 6.00)$/$\mathbf{3.92} (\pm 0.35)$} & \multicolumn{1}{c|}{$\mathbf{0.66} (\pm 5.00)$/$\mathbf{6.33} (\pm 0.85)$} 
\\
\cline{3-5}
\multicolumn{1}{|l|}{} &
\multicolumn{1}{l|}{LightGCN} & 
\multicolumn{1}{l|}{{\sc PersonalRank}} & \multicolumn{1}{c|}{$0.76 (\pm 4.00)$/$5.22 (\pm 0.29)$} & \multicolumn{1}{c|}{$0.56 (\pm 2.00)$/$7.06 (\pm 0.43)$}
\\
\cline{3-5}
\multicolumn{1}{|l|}{} &
\multicolumn{1}{l|}{} & 
\multicolumn{1}{l|}{{\sc Random}} &
\multicolumn{1}{c|}{$0.13 (\pm 3.00)$/$5.23 (\pm 1.60)$} & \multicolumn{1}{c|}{$0.09 (\pm 5.00)$/$8.70 (\pm 0.43)$}
\\
\cline{2-5}
\multicolumn{1}{|l|}{\textit{LastFM}} & 
\multicolumn{1}{l|}{} & 
\multicolumn{1}{l|}{GREASE} & 
\multicolumn{1}{c|}{$\mathbf{0.86} (\pm 7.00)$/$\mathbf{3.76} (\pm 0.51)$} & \multicolumn{1}{c|}{$\mathbf{0.68} (\pm 9.00)$/$\mathbf{6.30} (\pm 0.87)$}
\\
\cline{3-5}
\multicolumn{1}{|l|}{} &
\multicolumn{1}{l|}{NGCF} & 
\multicolumn{1}{l|}{{\sc PersonalRank}} & \multicolumn{1}{c|}{$0.80 (\pm 5.00)$/$5.10 (\pm 0.42)$} & \multicolumn{1}{c|}{$0.58 (\pm 3.00)$/$7.53 (\pm 0.42)$}
\\
\cline{3-5}
\multicolumn{1}{|l|}{} &
\multicolumn{1}{l|}{} & 
\multicolumn{1}{l|}{{\sc Random}} &
\multicolumn{1}{c|}{$0.19 (\pm 5.00)$/$5.77 (\pm 1.82)$} & \multicolumn{1}{c|}{$0.15 (\pm 4.00)$/$8.93 (\pm 0.47)$} 
\\
\hline
\hline
\multicolumn{1}{|l|}{} & 
\multicolumn{1}{l|}{} & 
\multicolumn{1}{l|}{GREASE} &
\multicolumn{1}{c|}{$\mathbf{0.86} (\pm 5.00)$/$\mathbf{3.76} (\pm 0.51)$} & \multicolumn{1}{c|}{$\mathbf{0.63} (\pm 3.65)$/$\mathbf{7.25} (\pm 0.62)$}
\\
\cline{3-5}
\multicolumn{1}{|l|}{} & 
\multicolumn{1}{l|}{LightGCN} & 
\multicolumn{1}{l|}{{\sc PersonalRank}} & \multicolumn{1}{c|}{$0.80 (\pm 7.00)$/$5.10 (\pm 0.42)$} & \multicolumn{1}{c|}{$0.56 (\pm 3.38)$/$7.36 (\pm 0.63)$}
\\
\cline{3-5}
\multicolumn{1}{|l|}{} & 
\multicolumn{1}{l|}{} & 
\multicolumn{1}{l|}{{\sc Random}} & 
\multicolumn{1}{c|}{$0.18 (\pm 6.00)$/$5.77 (\pm 1.82)$} & \multicolumn{1}{c|}{$0.09 (\pm 2.60)$/$7.67 (\pm 0.17)$} 
\\
\cline{2-5}
\multicolumn{1}{|l|}{\textit{Yelp}} & 
\multicolumn{1}{l|}{} & 
\multicolumn{1}{l|}{GREASE} &
\multicolumn{1}{c|}{$\mathbf{0.85} (\pm 7.00)$/$\mathbf{6.30} (\pm 0.53)$} & \multicolumn{1}{c|}{$\mathbf{0.68} (\pm 9.00)$/$\mathbf{6.30} (\pm 0.87)$} 
\\
\cline{3-5}
\multicolumn{1}{|l|}{} & 
\multicolumn{1}{l|}{NGCF} & 
\multicolumn{1}{l|}{{\sc PersonalRank}} & \multicolumn{1}{c|}{$0.79 (\pm 5.00)$/$6.79 (\pm 0.47)$} & \multicolumn{1}{c|}{$0.58 (\pm 3.00)$/$7.52 (\pm 0.42)$} 
\\
\cline{3-5}
\multicolumn{1}{|l|}{} & 
\multicolumn{1}{l|}{} & 
\multicolumn{1}{l|}{{\sc Random}} & 
\multicolumn{1}{c|}{$0.18 (\pm 2.00)$/$8.10 (\pm 0.52)$} & \multicolumn{1}{c|}{$0.15 (\pm 5.00)$/$8.93 (\pm 0.47)$}
\\
\hline
\end{tabular}
}
\caption{\label{tab:top-15-results} 
PN, PS, and EC of factual and counterfactual explanations generated by GREASE, {\sc PersonalRank}, and {\sc Random} for top-$15$ recommended items.}
\end{table*}

\begin{table*}[ht]
\scalebox{0.85}{
\centering
\begin{tabular}{lll|cc|}
\cline{4-5}
& & & \multicolumn{2}{c|}{{\bf Top-$20$ recommendations}}
\\
\hline
\multicolumn{1}{|l|}{\bf Dataset} & \multicolumn{1}{l|}{\thead{Recommendation\\Model}} & \multicolumn{1}{l|}{\thead{Explanation\\Method}} & \multicolumn{1}{c|}{\bf PN($\uparrow$)/EC($\downarrow$)} & \multicolumn{1}{c|}{\bf PS($\uparrow$)/EC($\downarrow$)}
\\
\hline
\multicolumn{1}{|l|}{} & 
\multicolumn{1}{l|}{} & 
\multicolumn{1}{l|}{GREASE} & 
\multicolumn{1}{c|}{$\mathbf{0.87} (\pm 7.00)$/$\mathbf{4.36} (\pm 0.46)$} & \multicolumn{1}{c|}{$\mathbf{0.65} (\pm 9.00)$/$\mathbf{6.35} (\pm 0.87)$} 
\\
\cline{3-5}
\multicolumn{1}{|l|}{} &
\multicolumn{1}{l|}{LightGCN} & 
\multicolumn{1}{l|}{{\sc PersonalRank}} & \multicolumn{1}{c|}{$0.76 (\pm 5.00)$/$5.79 (\pm 0.32)$} & \multicolumn{1}{c|}{$0.54 (\pm 5.00)$/$7.31 (\pm 0.52)$}
\\
\cline{3-5}
\multicolumn{1}{|l|}{} &
\multicolumn{1}{l|}{} & 
\multicolumn{1}{l|}{{\sc Random}} &
\multicolumn{1}{c|}{$0.09 (\pm 4.20)$/$5.82 (\pm 1.90)$} & \multicolumn{1}{c|}{$0.06 (\pm 7.00)$/$9.10 (\pm 0.49)$} 
\\
\cline{2-5}
\multicolumn{1}{|l|}{\textit{LastFM}} & 
\multicolumn{1}{l|}{} & 
\multicolumn{1}{l|}{GREASE} & 
\multicolumn{1}{c|}{$\mathbf{0.84} (\pm 11.00)$/$\mathbf{3.98} (\pm 0.67)$} & \multicolumn{1}{c|}{$\mathbf{0.64} (\pm 10.00)$/$\mathbf{6.35} (\pm 0.89)$}
\\
\cline{3-5}
\multicolumn{1}{|l|}{} &
\multicolumn{1}{l|}{NGCF} & 
\multicolumn{1}{l|}{{\sc PersonalRank}} & \multicolumn{1}{c|}{$0.77 (\pm 11.00)$/$5.32 (\pm 0.59)$} & \multicolumn{1}{c|}{$0.53 (\pm 5.00)$/$7.93 (\pm 0.58)$}
\\
\cline{3-5}
\multicolumn{1}{|l|}{} &
\multicolumn{1}{l|}{} & 
\multicolumn{1}{l|}{{\sc Random}} &
\multicolumn{1}{c|}{$0.17 (\pm 6.00)$/$5.92 (\pm 1.98)$} & \multicolumn{1}{c|}{$0.14 (\pm 5.00)$/$9.60 (\pm 0.52)$} 
\\
\hline
\hline
\multicolumn{1}{|l|}{} & 
\multicolumn{1}{l|}{} & 
\multicolumn{1}{l|}{GREASE} &
\multicolumn{1}{c|}{$\mathbf{0.83} (\pm 5.11)$/$\mathbf{4.93} (\pm 0.37)$} & \multicolumn{1}{c|}{$\mathbf{0.61} (\pm 3.92)$/$\mathbf{7.38} (\pm 0.75)$}
\\
\cline{3-5}
\multicolumn{1}{|l|}{} & 
\multicolumn{1}{l|}{LightGCN} & 
\multicolumn{1}{l|}{{\sc PersonalRank}} & \multicolumn{1}{c|}{$0.77 (\pm 4.20)$/$5.38(\pm 0.42)$} & \multicolumn{1}{c|}{$0.53 (\pm 3.42)$/$7.52 (\pm 0.73)$}
\\
\cline{3-5}
\multicolumn{1}{|l|}{} & 
\multicolumn{1}{l|}{} & 
\multicolumn{1}{l|}{{\sc Random}} & 
\multicolumn{1}{c|}{$0.16 (\pm 5.60)$/$5.42 (\pm 1.65)$} & \multicolumn{1}{c|}{$0.07 (\pm 1.90)$/$8.02 (\pm 0.18)$} 
\\
\cline{2-5}
\multicolumn{1}{|l|}{\textit{Yelp}} & 
\multicolumn{1}{l|}{} & 
\multicolumn{1}{l|}{GREASE} &
\multicolumn{1}{c|}{$\mathbf{0.75} (\pm 8.00)$/$\mathbf{6.70} (\pm 0.56)$} & \multicolumn{1}{c|}{$\mathbf{0.63} (\pm 10.00)$/$\mathbf{6.37} (\pm 0.86)$} 
\\
\cline{3-5}
\multicolumn{1}{|l|}{} & 
\multicolumn{1}{l|}{NGCF} & 
\multicolumn{1}{l|}{{\sc PersonalRank}} & \multicolumn{1}{c|}{$0.71 (\pm 6.12)$/$7.12 (\pm 0.52)$} & \multicolumn{1}{c|}{$0.53 (\pm 5.00)$/$7.93 (\pm 0.58)$} 
\\
\cline{3-5}
\multicolumn{1}{|l|}{} & 
\multicolumn{1}{l|}{} & 
\multicolumn{1}{l|}{{\sc Random}} & 
\multicolumn{1}{c|}{$0.12 (\pm 2.00)$/$8.60 (\pm 0.55)$} & \multicolumn{1}{c|}{$0.14 (\pm 6.00)$/$9.50 (\pm 0.52)$}
\\
\hline
\end{tabular}
}
\caption{\label{tab:top-20-results} 
PN, PS, and EC of factual and counterfactual explanations generated by GREASE, {\sc PersonalRank}, and {\sc Random} for top-$20$ recommended items.}
\end{table*}

\section{Conclusion and Future Work}
\label{sec:conclusion}
In this work, we presented GREASE, an explanation method for GNN-based recommendations. 
GREASE operates in a fully ``black-box'' setting, i.e., it needs only to access the predicted ratings output by the GNN-based recommender system.
Using a common framework, it can generate factual and counterfactual explanations to capture the {\em sufficient} and {\em necessary} conditions for an item to be recommended to a user, respectively.
Experiments conducted on two real-world datasets combined with two well-known GNN-based recommender systems showed that GREASE outperforms baseline approaches.

Despite a lot of recent effort into explainable AI, we argue that explaining recommendations provided by AI/ML models (and, in particular, GNNs) is still underexplored. 
In future work, we plan to: {\em(i)} unify the framework introduced by GREASE to combine simultaneously factual and counterfactual explanations; {\em(ii)} extend our experiments to more datasets; and, possibly, {\em(iii)} adapt existing GNN explanation methods to the recommendation setting.
%
%
%
%

\end{document}